\documentclass[a4paper,11pt]{article}
\usepackage{pos}
\usepackage{blindtext}
\usepackage{etoolbox}
\usepackage{hyperref}
\usepackage{url}
\usepackage{xurl}
\usepackage{graphicx}
\usepackage{wrapfig}
\usepackage{microtype}
\usepackage{ragged2e} 
\usepackage{setspace}
\usepackage[utf8]{inputenc}  
\usepackage[T1]{fontenc}
\usepackage[left]{lineno}  
\usepackage{enumitem}

\title{Education and Outreach Across Seventeen Countries of the Pierre Auger Collaboration}
 \ShortTitle{Pierre Auger Collaboration Outreach around the world}

\author*[a, b]{Fabio Convenga}

\affiliation[a]{Gran Sasso Science Institute (GSSI), Via Iacobucci 2, I-67100 L’Aquila, Italy}
\affiliation[b]{Istituto Nazionale di Fisica Nucleare (INFN)–Laboratori Nazionali del Gran Sasso, I-67100 Assergi,
L’Aquila, Italy}

\onbehalf{for the Pierre Auger Collaboration$^c$} 
\affiliation[c]{Observatorio Pierre Auger, Av.\ San Mart{\'\i}n Norte 304, 5613 Malarg\"ue, Argentina\\
Full author list: {\rm\url{https://www.auger.org/archive/authors_icrc_2025.html}}}

\emailAdd{spokespersons@auger.org}

\abstract{The Pierre Auger Observatory, dedicated to measuring ultra-high energy cosmic rays, has been promoting for more than two decades educational and scientific outreach activities to make its results known in an understandable language to diverse audiences. 
Among its most notable efforts, we can mention the creation of a visitor center at the Observatory site in Malargüe, Argentina, the production of brochures, posters, videos, talks, and special actions with the community in the site of the Observatory and beyond. 
In addition to joint efforts, collaborators from participating countries carry out local efforts, some of them related to national initiatives in their respective regions, such as the International Cosmic Day, Masterclasses, exhibitions of artworks with the theme of astroparticles, the Night of the Stars, European Researchers' Night, summer schools and initiatives to improve the gender balance in the science community. 
In addition, there have been board games based on the dynamics of the observatory's work, online graphic viewers of the different events, talks, workshops, etc. 
In recent years, the Pierre Auger Outreach group has focused on presenting actions that directly impact the community in Malarg\"ue. 
However, this time, special emphasis will be placed on highlighting the outreach efforts of Pierre Auger collaborators in various countries.}

\ConferenceLogo{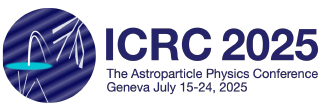}

\FullConference{%
39th International Cosmic Ray Conference (ICRC2025)\\
 15 – 24 July, 2025\\
  Geneva, Switzerland}

\begin{document}
\maketitle

\section{Introduction}
The Pierre Auger Observatory, located near Malarg\"ue in Argentina, is the world’s largest facility for studying ultra-high-energy cosmic rays. 
While its primary mission is scientific, the Pierre Auger Collaboration has consistently promoted science communication and education, making the physics of cosmic rays accessible to a wide audience through local, national, and international initiatives. 
These outreach efforts, carried out both at the Observatory site and by collaborators in many participating countries, have included public exhibitions, school visits, workshops, and participation in global events. 
In parallel, specific actions have aimed to address broader issues within the scientific community, such as promoting gender balance and increasing the participation of underrepresented groups in science.

Among these initiatives, the Auger International Masterclass \cite{masterclass} stands out as a key educational program, enabling high school students from around the world to analyze open data \cite{opendata} from the Observatory using interactive tools and learn about cosmic ray physics in a collaborative environment.

In addition, two initiatives that have had a strong impact on students and the local community are the Malargüe Science Fair and the Observatory’s Visitor Center. 
The Science Fair takes place every two years and involves students from different age groups, including primary schools and adult education, who prepare and present science projects. 
A selection of these projects is brought to Malargüe, where students can discuss their work with scientists from the Pierre Auger Collaboration. 

The Visitor Center, located at the Observatory site in Malargüe, is open throughout the year and offers an interactive exhibition for the public. 

At the beginning of 2023, the exhibition was updated with new elements, including a virtual assistant, small-scale models of the detectors, working equipment, and a virtual reality experience that simulates a balloon flight over the Observatory.

In addition to activities based in Malargüe, members of the Pierre Auger Collaboration have developed a wide range of outreach initiatives in their home countries. 
This report presents a summary of the educational and outreach activities carried out by the Collaboration in recent years around the world. 
In addition, the third section focuses on actions that promote inclusion, accessibility, and gender balance in science. The following section highlights the use of social media by the Collaboration.

\section{Education, Outreach and Public Engagement Around the World}
The Pierre Auger Collaboration includes more than 500 scientists from 17 countries across all continents, many of whom are actively involved in educational and outreach activities within their local communities.
In addition to these local efforts, all members of the Collaboration contribute to the organization and coordination of major international outreach initiatives, including the International Cosmic Day (ICD), the European Researchers’ Night (ERN), and the Auger International Masterclass.
This section presents a selection of these initiatives.

\begin{figure}[ht]
  \centering
  \includegraphics[width=0.8\textwidth]{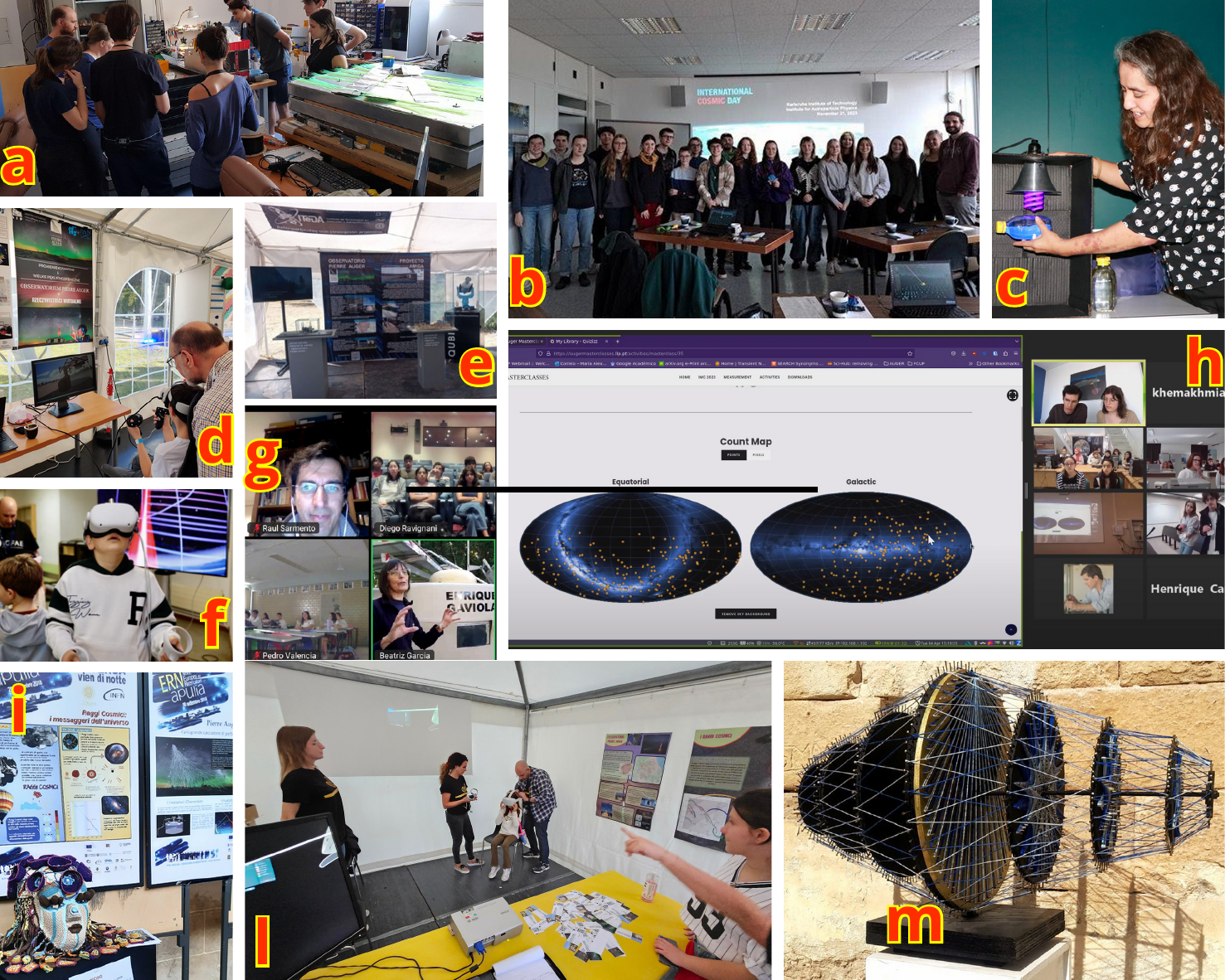}
  \caption{%
    Selection of outreach activities carried out by members of the Pierre Auger Collaboration in different countries.
    (\textbf{a}) Students visiting Institute of Space Science (ISS) laboratories. 
    (\textbf{b}) Group photo during the International Cosmic Day 2023 at Karlsruher Institut für Technologie (KIT). 
    (\textbf{c}) Demonstration of fluorescent light emission. 
    (\textbf{d}) LIP VR system used during ERN 2023 in Poland. 
    (\textbf{e}) Instituto de Tecnologías en Detección y Astropartículas (ITeDA) stand during \textit{Mendociencia}. 
    (\textbf{f}) Students exploring EAS using a VR headset in Spain. 
    (\textbf{g}) Online visit of the Observatory site during the Auger International Masterclass. 
    (\textbf{h}) Screenshot from a remote Auger Masterclass session. 
    (\textbf{i}) Posters and artistic materials from ERN 2024 in Lecce. 
    (\textbf{l}) Students using the VR system in L'Aquila during \textit{Street Science} 2024. 
    (\textbf{m}) Sculpture inspired by astroparticle physics, exhibited during ERN 2024 in Lecce.%
  }
  \label{fig:outreach-collage}
\end{figure}

In Argentina, activities related to the Observatory have been featured in events such as \textit{Argentine Science Week} 2023 and \textit{Mendociencia}, a science fair organized by the Faculty of Basic and Natural Sciences of the University of Cuyo. 
In this context, the Observatory was presented at a dedicated stand with the use of reproductions of its detectors.
Additionally, several meetings have been organized—and continue to be held—by the Auger Collaboration community in Argentina to present cosmic ray physics and the Observatory to both university and high school students.

In Australia, collaborators developed educational materials for schools and universities to introduce an educational muon detection system called “EduMOD” \cite{edumod}.
To support this initiative, presentations have been organized for high school teachers nationwide. 
Additionally, Australia Collaboration groups have been actively giving public lectures to science-interested groups, such as Rotary clubs, highlighting the physics research conducted by the Collaboration.

In Belgium, the Brussels group contributed to the creation of a video \cite{videobelgium} for the 50th anniversary of the Interuniversity Institute for High Energies (IIHE), presented during a colloquium held in September 2022. 
The video includes a segment on the Pierre Auger Observatory, highlighting its scientific relevance. 

In addition, it is worth mentioning the educational activities carried out in Brazil, aimed at engaging students in astroparticle physics. 
Among these, a public lecture was held in 2021 at the Instituto Federal do Acre, located in the middle of the Amazon rainforest, focusing on neutrinos and ultra-high-energy cosmic rays.
The event introduced the Observatory and related experiments to students and the wider public.

In Colombia, the outreach program of the Collaboration local groups engages the public through interactive events that demonstrate how cosmic rays impact daily life. 
One example is the regular shows at the Universidad Industrial de Santander planetarium \cite{planetarium}, where visitors learn about the relationship between cosmic rays and everyday technologies such as GPS and medical imaging. 

In Czech Republic, outreach activities linked to the Pierre Auger Observatory took place during the 11th IDPASC School in Olomouc (August 29 – September 7, 2023). 
The program included interviews \cite{czechinterviews} with leading scientists such as Francis Halzen, Ralph Engel, and Mario Pimenta. 
The event was supported by the infrastructure national projects AUGER-CZ and CTA-CZ.

In France, in June 2023, a seminar titled “Les rayons cosmiques et la traque des plus énergétiques” \cite{corinne} was presented to the Alps section of the  Société Française de Physique. 
The talk focused on the Observatory and its main results and was attended by physicists from various fields, retired researchers, and students. 
On March 26, 2024, a visit was held at Lycée Élie Cartan in La Tour-du-Pin (Isère), where students were introduced to cosmic rays, their detection methods, and the activities of the Observatory.

In Germany, members of the Collaboration carried out a variety of educational activities across several institutions. 
On November 21, 2023, as part of International Cosmic Day (ICD), 17 high school students from eight different schools participated in an introductory presentation on cosmic rays, followed by group activities in which they took measurements of muons. 
The students also took part in the Auger Masterclass. 
The day ended with an international video call, during which participants shared their results with students from other institutions, including schools in the UK and China and the University of Würzburg. 
Another activity held by German collaborators was the \textit{Girl’s Day Mädchen-Zukunftstag} event in April 2024, with the activity “Astroteilchenphysik - Mit kleinsten Teilchen das Universum entdecken (Astroparticle Physics- Discover the Universe with the smallest particles)”, where 14-year-old girls learned about astroparticle physics through the Observatory, they built a simple cosmic ray detector, and performed basic data analysis.

In Italy, members of the Collaboration have organized numerous awareness-raising and training activities through the various sections of the Istituto Nazionale di Fisica Nucleare (INFN) and Italian universities. 
As part of these activities, students explored various aspects of cosmic rays research, carried out basic measurements, visited laboratories, and interacted directly with researchers. 
Among these efforts, in Lecce the local group took part in ERN 2024 by participating in an exhibition held in the city center, presenting several educational activities, including \textit{Creating by Imagining – Science Represented through Art}. 
The activity featured sculptures created by students from the Liceo Artistico Ciardo Pellegrino, inspired by astroparticle science and developed within the national project “Art \& Science Across Italy" \cite{Galati2024}. 
Moreover, the group in L’Aquila participated in the local event “Street Science” organized in conjunction with ERN 2024, offering an activity open to students and visitors of all ages. 
The experience included a virtual tour of the Pierre Auger Observatory, made possible through the 3D virtual reality system developed by the Portugal collaborators. 
In Palermo, joining this national effort, local group members took part in the “Physics \& Drinks” event \cite{palermo} in March 2025, where they introduced the Observatory to university and high school students in an informal setting.

In Mexico, members of the Collaboration have conducted several educational and outreach initiatives, including conferences, workshops, and hands-on activities aimed at students ranging from secondary school to university level.
During the \textit{Ciencia Recreativa} event held from the 14th to the 18th of August 2023, the Puebla de Zaragoza Auger group taught one of the courses for middle and high school students. 
In this course, participants were introduced to the physics behind the research conducted at the Observatory through the use of videos, simulations, and interactive demonstrations. 

\begin{figure}[h]
    \centering
    \begin{minipage}[b]{0.4\textwidth}
        \centering
        \includegraphics[width=\textwidth]{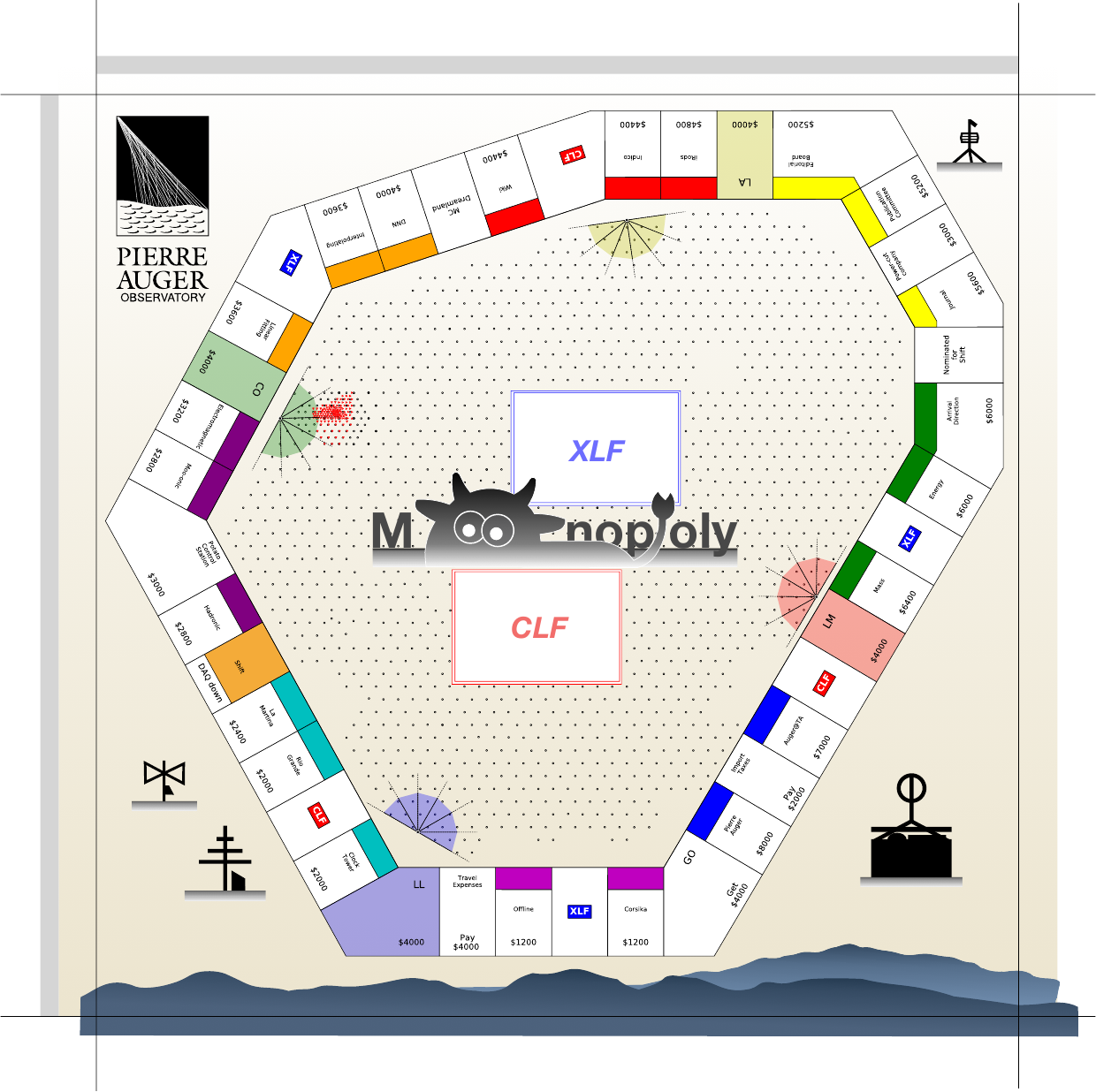}
    \end{minipage}
    \begin{minipage}[b]{0.4\textwidth}
        \centering
        \includegraphics[width=\textwidth]{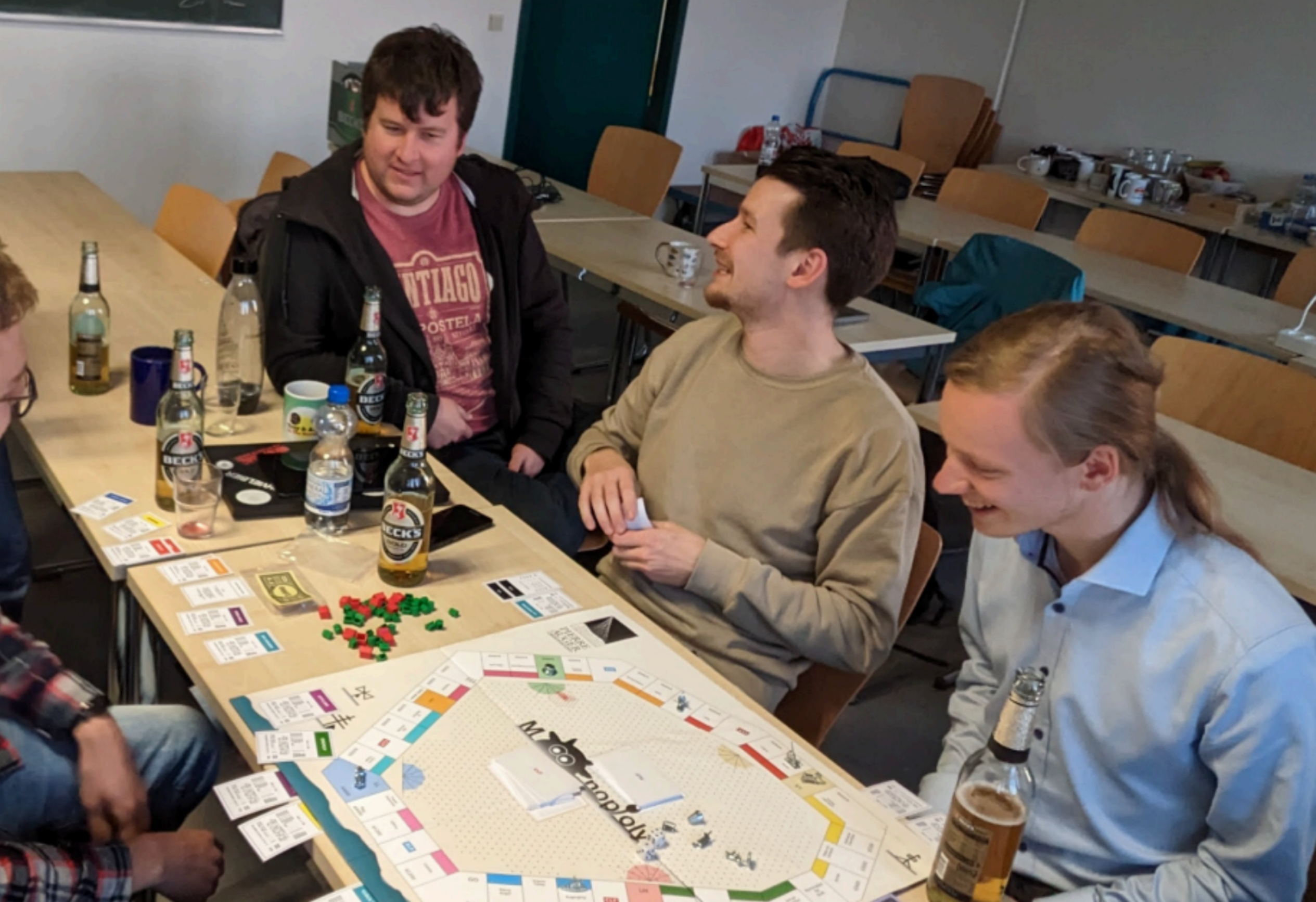}     
    \end{minipage}
    \caption{Customized Auger-themed Monopoly game. (a) Game board layout. (b) Game in use during a social occasion.}
    \label{fig:monopoly}
\end{figure}

In the Netherlands, the Nijmegen local group created a Pierre Auger Observatory–themed Monopoly game as a playful initiative to raise awareness.
Featuring elements of the Observatory and Collaboration-inspired content, the board was brought to Malargüe and used to engage participants in a fun and informal way. An example of the game is shown in Figure~\ref{fig:monopoly}.
Sometimes, the best way to learn about cosmic rays is to try to buy the Coihueco Fluorescence Detector site \cite{fd} before your opponents do!

In Poland, the group at the Institute of Nuclear Physics Polish Academy of Science (IFJ PAN) in Kraków produced an educational film titled \textit{Physics is simple – Time dilation} \cite{polandfilm}, which explains the concept of time dilation using the example of atmospheric muons and their detection at the Pierre Auger Observatory. 
In addition, in September 2023, the group used the LIP 3D virtual reality system to present the Observatory during the ERN in Kraków. 
The activity was well received, attracting participants beyond the official duration of the event.

In Portugal, collaborators are responsible for maintaining the Auger International Masterclass and its web infrastructure, while content development and session delivery involve contributions from the entire collaboration.
An important contribution from Portuguese collaborators is the 3D virtual reality (VR) system developed by the Laboratory of Instrumentation and Experimental Particle Physics (LIP). 
The system allows users to virtually explore how extensive air showers (EAS), generated by cosmic rays in the atmosphere, interact with the 1,600 surface detector stations that compose the array of the Pierre Auger Observatory. In addition, users can examine a single station in detail, walking around and even inside the structure, observing the photomultipliers in action as they detect the Cherenkov light produced by EAS particles passing through the detector. In early 2024, the VR system was expanded with new content based on drone footage of the Observatory and improved with additional viewing stations to increase accessibility. 

In Romania, in May 2023, a group of local science teachers visited the Bucharest collaborators to learn about the physics studied at the Pierre Auger Observatory. 
During the same period, an Auger-related talk was presented during the \textit{Astrofest} 2023 \cite{astrofest} public event in Bucharest. In September 2024, a group of eight high school students visited the Bucharest Collaboration group for a week-long program focused on astroparticle physics. 
The initiative was proposed by the students themselves following their participation, in March 2024, in the Auger International Masterclass. They explored Auger open data, reproduced plots, visited the Horia Hulubei National Institute of Physics and Nuclear Engineering (IFIN HH) group for practical work with detectors, and completed with individual mini-projects focused on the detection of cosmic rays and related data analysis.

In Spain, since the beginning of 2023, the 3D VR system developed by LIP has been used to offer immersive experiences of the Observatory to visiting students and the public. 
During the 2023/2024 school year, more than 200 students from the Santiago di Compostela region, participated in guided learning sessions that integrated the VR experience. The activity was also presented during a public open day held in November 2023, which welcomed over 200 visitors of all ages.
The initiative also received media attention, including the program \textit{Atmosféricos} on Televisión de Galicia \cite{atmosfericos}, and expanded its visibility by reaching a wider audience through social media.

Finally, a particular mention should be made of the contribution from the United States. 
For example, each year, a grant is provided to support a secondary school student from Malargüe in pursuing academic studies at Michigan Technological University. 
Furthermore, several of the interactive devices and exhibits at the Visitor Center were donated by U.S. institutions, significantly enriching the educational experience for all visitors.

\section{Equity, Inclusion and Accessibility}
The Pierre Auger Collaboration dedicates part of its commitment to activities improving equity, inclusion, and accessibility through a series of initiatives aimed at involving underrepresented groups in the field of science. 
These include exhibitions and artistic activities, workshops focusing on accessibility, and collaborative initiatives with schools and local communities.

\subsection{Women and Girls in Science}
In 2024, as part of this effort and consistent with its regular engagement, the Collaboration marked two significant occasions: the International Day of Women and Girls in Science (February 11) and International Women’s Day (March 8). 
To honor these occasions, two complementary initiatives were launched, one dedicated to women scientists within the Collaboration, and the other addressed to the general public.

Women collaborators were invited to contribute with personal reflections, describing their daily lives, how they balance scientific work with personal commitments, and sharing other passions or interests beyond their professional roles, such as music, sports, or creative hobbies. 
The material collected was later used to update a Observatory website page \cite{womenscience} dedicated to highlighting the role of women in the Collaboration. 

For the public, an art and science contest entitled \textit{Women and Girls in Science 2023} was announced, inviting participants to submit drawings focusing on the role of women and girls in science, aligned with the 2023 UNESCO motto “Innovate. Demonstrate. Raise. Advance. Sustain (I.D.E.A.S.)”. 
The selected artworks have illustrated the months of the Pierre Auger Observatory 2024 calendar, including one image featured on the cover. 
This initiative aimed to promote creative engagement with science while highlighting the presence of women and girls in scientific contexts.

The selected artworks were also part of a touring exhibition showcased at several locations throughout Argentina in 2024. 
The exhibition opened at the Planetarium in Malargüe in February. 
It was later displayed at the Malargüe Convention Center in March, with support from the local Women’s Affairs Office, followed by installations at Arizu Space in Godoy Cruz \cite{arizuspace} in April, the National High School-UNL in La Plata in September, and the National Technological University in Mendoza (UTN) in October - November.

\subsection{Accessibility and Engagement with Diverse Audiences}
\textit{The Observatory Goes to the School} is an outreach program that brings science and the work of the Pierre Auger Observatory to students and teachers, especially in the Malarg\"ue region. 
As part of this initiative, staff members have visited both rural and urban schools, reaching around 5000 students in just a few months. 
Meetings and hands-on demonstrations have been organized, and educational material has been shared in the Malarg\"ue area and nearby communities such as El Manzano and El Alambrado, located about 100 km and 110 km from the town, respectively.

\begin{figure}[h]
    \centering
\includegraphics[width=0.75\textwidth]{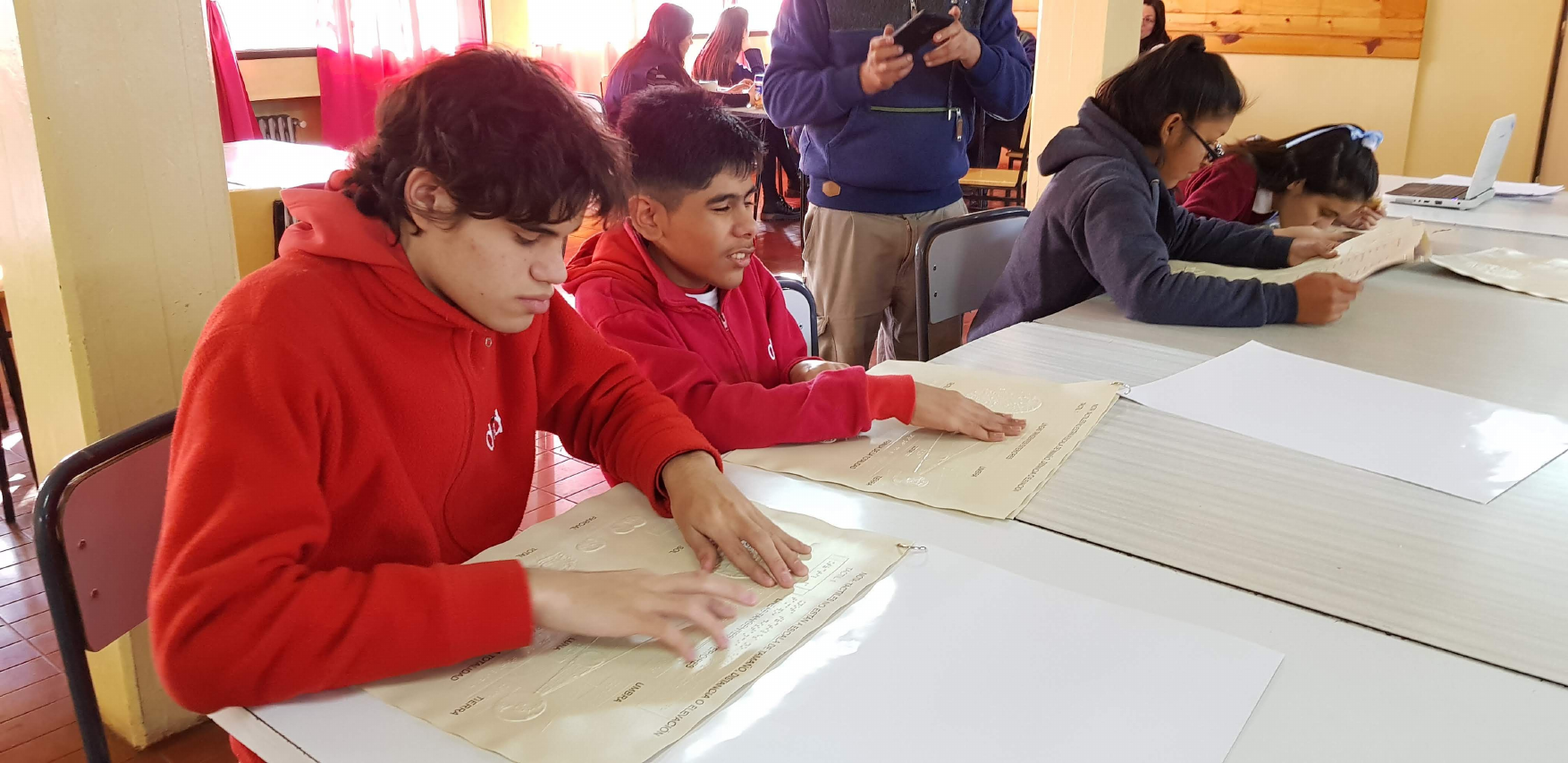}
    \caption{Students with visual impairments at the Helen Keller School in Godoy Cruz (Mendoza) explore tactile science materials.}
    \label{fig:tactile_materials}
\end{figure}

The program also includes activities focused on accessibility and inclusion. A special activity was organized at the Maurin Navarro School to engage blind, deaf, and neurodiverse students. 
These efforts are in line with the objectives of the International Astronomical Union (IAU) Working Group for Equity and Inclusion\cite{iauworkinggroup}, which promotes access to astronomy for individuals of all backgrounds and abilities. To foster long-term engagement, a series of workshops was launched involving both students and teachers. 
In parallel, local collaborations are being explored, such as with technical schools in Malargüe for the development of tactile materials using 3D printing. An example of the use of these materials is shown in Figure~\ref{fig:tactile_materials}.

In the same spirit of inclusion, on December 5, 2024, the Observatory marked the International Day of Persons with Disabilities with a dedicated workshop held at UTN in Mendoza. During the event, the software tool \textit{sonoUno} \cite{Casado1} was presented, a platform developed to sonify scientific data, enabling auditory data representation for visually impaired users. 
 
The training involved both blind and sighted participants, including students, educators, and musicians. 
Activities featured tactile models, an introduction to Braille, and hands-on use of tools such as Geiger counters, actuators, and transducers. 

\section{Social Media}
The Pierre Auger Collaboration actively informs the public via social media platforms such as Instagram, Blue Sky, and Facebook. 
Weekly posts highlight scientific news, outreach efforts, and behind-the-scenes insights into the Observatory. 
The maintenance and development of these communication channels are made possible thanks to the cooperation of members of the Collaboration.

In addition, \textit{Auger en Foco} is published monthly, presenting highlights of the studied physics, including new research, outreach initiatives, and social moments. 
Each issue is shared monthly on Instagram, and the complete collection is available at \cite{augerenfoco}.

\apptocmd{\thebibliography}{\small\setlength{\itemsep}{0pt}}{}{}

\newpage

\section*{The Pierre Auger Collaboration}

{\footnotesize\setlength{\baselineskip}{10pt}
\noindent
\begin{wrapfigure}[11]{l}{0.12\linewidth}
\vspace{-4pt}
\includegraphics[width=0.98\linewidth]{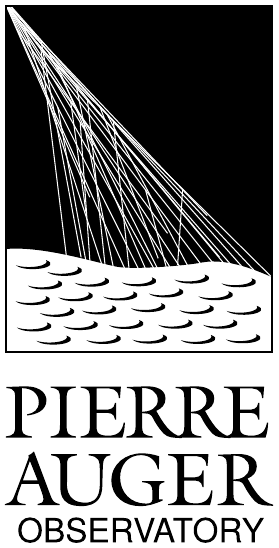}
\end{wrapfigure}
\begin{sloppypar}\noindent
A.~Abdul Halim$^{13}$,
P.~Abreu$^{70}$,
M.~Aglietta$^{53,51}$,
I.~Allekotte$^{1}$,
K.~Almeida Cheminant$^{78,77}$,
A.~Almela$^{7,12}$,
R.~Aloisio$^{44,45}$,
J.~Alvarez-Mu\~niz$^{76}$,
A.~Ambrosone$^{44}$,
J.~Ammerman Yebra$^{76}$
% ...

\end{sloppypar}

\newlength{\decorativeline}
\setlength{\decorativeline}{0.5\linewidth}
\addtolength{\decorativeline}{-2.9em}

\begin{center}
\raisebox{0.2ex}{\rule{0.1\linewidth}{0.4pt}}%
\hspace{0.6ex}\raisebox{-0.1ex}{$\bullet$}\hspace{0.6ex}%
\raisebox{0.2ex}{\rule{0.1\linewidth}{0.4pt}}
\end{center}

\vspace{1ex}
\begin{description}[labelsep=0.2em,align=right,labelwidth=0.7em,labelindent=0em,leftmargin=2em,noitemsep,before={\renewcommand\makelabel[1]{##1 }}]
\item[$^{1}$] Centro At\'omico Bariloche and Instituto Balseiro (CNEA-UNCuyo-CONICET), San Carlos de Bariloche, Argentina
\item[$^{2}$] Departamento de F\'\i{}sica and Departamento de Ciencias de la Atm\'osfera y los Oc\'eanos, FCEyN, Universidad de Buenos Aires and CONICET, Buenos Aires, Argentina
\item[$^{3}$] IFLP, Universidad Nacional de La Plata and CONICET, La Plata, Argentina
\item[$^{4}$] Instituto de Astronom\'\i{}a y F\'\i{}sica del Espacio (IAFE, CONICET-UBA), Buenos Aires, Argentina
\item[$^{5}$] Instituto de F\'\i{}sica de Rosario (IFIR) -- CONICET/U.N.R.\ and Facultad de Ciencias Bioqu\'\i{}micas y Farmac\'euticas U.N.R., Rosario, Argentina
\item[$^{6}$] Instituto de Tecnolog\'\i{}as en Detecci\'on y Astropart\'\i{}culas (CNEA, CONICET, UNSAM), and Universidad Tecnol\'ogica Nacional -- Facultad Regional Mendoza (CONICET/CNEA), Mendoza, Argentina
\item[$^{7}$] Instituto de Tecnolog\'\i{}as en Detecci\'on y Astropart\'\i{}culas (CNEA, CONICET, UNSAM), Buenos Aires, Argentina
\item[$^{8}$] International Center of Advanced Studies and Instituto de Ciencias F\'\i{}sicas, ECyT-UNSAM and CONICET, Campus Miguelete -- San Mart\'\i{}n, Buenos Aires, Argentina
\item[$^{9}$] Laboratorio Atm\'osfera -- Departamento de Investigaciones en L\'aseres y sus Aplicaciones -- UNIDEF (CITEDEF-CONICET), Argentina
\item[$^{10}$] Observatorio Pierre Auger, Malarg\"ue, Argentina
\item[$^{11}$] Observatorio Pierre Auger and Comisi\'on Nacional de Energ\'\i{}a At\'omica, Malarg\"ue, Argentina
\item[$^{12}$] Universidad Tecnol\'ogica Nacional -- Facultad Regional Buenos Aires, Buenos Aires, Argentina
\item[$^{13}$] University of Adelaide, Adelaide, S.A., Australia
\item[$^{14}$] Universit\'e Libre de Bruxelles (ULB), Brussels, Belgium
\item[$^{15}$] Vrije Universiteit Brussels, Brussels, Belgium
\item[$^{16}$] Centro Brasileiro de Pesquisas Fisicas, Rio de Janeiro, RJ, Brazil
\item[$^{17}$] Centro Federal de Educa\c{c}\~ao Tecnol\'ogica Celso Suckow da Fonseca, Petropolis, Brazil
\item[$^{18}$] Instituto Federal de Educa\c{c}\~ao, Ci\^encia e Tecnologia do Rio de Janeiro (IFRJ), Brazil
\item[$^{19}$] Universidade de S\~ao Paulo, Escola de Engenharia de Lorena, Lorena, SP, Brazil
\item[$^{20}$] Universidade de S\~ao Paulo, Instituto de F\'\i{}sica de S\~ao Carlos, S\~ao Carlos, SP, Brazil
\item[$^{21}$] Universidade de S\~ao Paulo, Instituto de F\'\i{}sica, S\~ao Paulo, SP, Brazil
\item[$^{22}$] Universidade Estadual de Campinas (UNICAMP), IFGW, Campinas, SP, Brazil
\item[$^{23}$] Universidade Estadual de Feira de Santana, Feira de Santana, Brazil
\item[$^{24}$] Universidade Federal de Campina Grande, Centro de Ciencias e Tecnologia, Campina Grande, Brazil
\item[$^{25}$] Universidade Federal do ABC, Santo Andr\'e, SP, Brazil
\item[$^{26}$] Universidade Federal do Paran\'a, Setor Palotina, Palotina, Brazil
\item[$^{27}$] Universidade Federal do Rio de Janeiro, Instituto de F\'\i{}sica, Rio de Janeiro, RJ, Brazil
\item[$^{28}$] Universidad de Medell\'\i{}n, Medell\'\i{}n, Colombia
\item[$^{29}$] Universidad Industrial de Santander, Bucaramanga, Colombia
\item[$^{30}$] Charles University, Faculty of Mathematics and Physics, Institute of Particle and Nuclear Physics, Prague, Czech Republic
\item[$^{31}$] Institute of Physics of the Czech Academy of Sciences, Prague, Czech Republic
\item[$^{32}$] Palacky University, Olomouc, Czech Republic
\item[$^{33}$] CNRS/IN2P3, IJCLab, Universit\'e Paris-Saclay, Orsay, France
\item[$^{34}$] Laboratoire de Physique Nucl\'eaire et de Hautes Energies (LPNHE), Sorbonne Universit\'e, Universit\'e de Paris, CNRS-IN2P3, Paris, France
\item[$^{35}$] Univ.\ Grenoble Alpes, CNRS, Grenoble Institute of Engineering Univ.\ Grenoble Alpes, LPSC-IN2P3, 38000 Grenoble, France
\item[$^{36}$] Universit\'e Paris-Saclay, CNRS/IN2P3, IJCLab, Orsay, France
\item[$^{37}$] Bergische Universit\"at Wuppertal, Department of Physics, Wuppertal, Germany
\item[$^{38}$] Karlsruhe Institute of Technology (KIT), Institute for Experimental Particle Physics, Karlsruhe, Germany
\item[$^{39}$] Karlsruhe Institute of Technology (KIT), Institut f\"ur Prozessdatenverarbeitung und Elektronik, Karlsruhe, Germany
\item[$^{40}$] Karlsruhe Institute of Technology (KIT), Institute for Astroparticle Physics, Karlsruhe, Germany
\item[$^{41}$] RWTH Aachen University, III.\ Physikalisches Institut A, Aachen, Germany
\item[$^{42}$] Universit\"at Hamburg, II.\ Institut f\"ur Theoretische Physik, Hamburg, Germany
\item[$^{43}$] Universit\"at Siegen, Department Physik -- Experimentelle Teilchenphysik, Siegen, Germany
\item[$^{44}$] Gran Sasso Science Institute, L'Aquila, Italy
\item[$^{45}$] INFN Laboratori Nazionali del Gran Sasso, Assergi (L'Aquila), Italy
\item[$^{46}$] INFN, Sezione di Catania, Catania, Italy
\item[$^{47}$] INFN, Sezione di Lecce, Lecce, Italy
\item[$^{48}$] INFN, Sezione di Milano, Milano, Italy
\item[$^{49}$] INFN, Sezione di Napoli, Napoli, Italy
\item[$^{50}$] INFN, Sezione di Roma ``Tor Vergata'', Roma, Italy
\item[$^{51}$] INFN, Sezione di Torino, Torino, Italy
\item[$^{52}$] Istituto di Astrofisica Spaziale e Fisica Cosmica di Palermo (INAF), Palermo, Italy
\item[$^{53}$] Osservatorio Astrofisico di Torino (INAF), Torino, Italy
\item[$^{54}$] Politecnico di Milano, Dipartimento di Scienze e Tecnologie Aerospaziali , Milano, Italy
\item[$^{55}$] Universit\`a del Salento, Dipartimento di Matematica e Fisica ``E.\ De Giorgi'', Lecce, Italy
\item[$^{56}$] Universit\`a dell'Aquila, Dipartimento di Scienze Fisiche e Chimiche, L'Aquila, Italy
\item[$^{57}$] Universit\`a di Catania, Dipartimento di Fisica e Astronomia ``Ettore Majorana``, Catania, Italy
\item[$^{58}$] Universit\`a di Milano, Dipartimento di Fisica, Milano, Italy
\item[$^{59}$] Universit\`a di Napoli ``Federico II'', Dipartimento di Fisica ``Ettore Pancini'', Napoli, Italy
\item[$^{60}$] Universit\`a di Palermo, Dipartimento di Fisica e Chimica ''E.\ Segr\`e'', Palermo, Italy
\item[$^{61}$] Universit\`a di Roma ``Tor Vergata'', Dipartimento di Fisica, Roma, Italy
\item[$^{62}$] Universit\`a Torino, Dipartimento di Fisica, Torino, Italy
\item[$^{63}$] Benem\'erita Universidad Aut\'onoma de Puebla, Puebla, M\'exico
\item[$^{64}$] Unidad Profesional Interdisciplinaria en Ingenier\'\i{}a y Tecnolog\'\i{}as Avanzadas del Instituto Polit\'ecnico Nacional (UPIITA-IPN), M\'exico, D.F., M\'exico
\item[$^{65}$] Universidad Aut\'onoma de Chiapas, Tuxtla Guti\'errez, Chiapas, M\'exico
\item[$^{66}$] Universidad Michoacana de San Nicol\'as de Hidalgo, Morelia, Michoac\'an, M\'exico
\item[$^{67}$] Universidad Nacional Aut\'onoma de M\'exico, M\'exico, D.F., M\'exico
\item[$^{68}$] Institute of Nuclear Physics PAN, Krakow, Poland
\item[$^{69}$] University of \L{}\'od\'z, Faculty of High-Energy Astrophysics,\L{}\'od\'z, Poland
\item[$^{70}$] Laborat\'orio de Instrumenta\c{c}\~ao e F\'\i{}sica Experimental de Part\'\i{}culas -- LIP and Instituto Superior T\'ecnico -- IST, Universidade de Lisboa -- UL, Lisboa, Portugal
\item[$^{71}$] ``Horia Hulubei'' National Institute for Physics and Nuclear Engineering, Bucharest-Magurele, Romania
\item[$^{72}$] Institute of Space Science, Bucharest-Magurele, Romania
\item[$^{73}$] Center for Astrophysics and Cosmology (CAC), University of Nova Gorica, Nova Gorica, Slovenia
\item[$^{74}$] Experimental Particle Physics Department, J.\ Stefan Institute, Ljubljana, Slovenia
\item[$^{75}$] Universidad de Granada and C.A.F.P.E., Granada, Spain
\item[$^{76}$] Instituto Galego de F\'\i{}sica de Altas Enerx\'\i{}as (IGFAE), Universidade de Santiago de Compostela, Santiago de Compostela, Spain
\item[$^{77}$] IMAPP, Radboud University Nijmegen, Nijmegen, The Netherlands
\item[$^{78}$] Nationaal Instituut voor Kernfysica en Hoge Energie Fysica (NIKHEF), Science Park, Amsterdam, The Netherlands
\item[$^{79}$] Stichting Astronomisch Onderzoek in Nederland (ASTRON), Dwingeloo, The Netherlands
\item[$^{80}$] Universiteit van Amsterdam, Faculty of Science, Amsterdam, The Netherlands
\item[$^{81}$] Case Western Reserve University, Cleveland, OH, USA
\item[$^{82}$] Colorado School of Mines, Golden, CO, USA
\item[$^{83}$] Department of Physics and Astronomy, Lehman College, City University of New York, Bronx, NY, USA
\item[$^{84}$] Michigan Technological University, Houghton, MI, USA
\item[$^{85}$] New York University, New York, NY, USA
\item[$^{86}$] University of Chicago, Enrico Fermi Institute, Chicago, IL, USA
\item[$^{87}$] University of Delaware, Department of Physics and Astronomy, Bartol Research Institute, Newark, DE, USA
\item[] -----
\item[$^{a}$] Max-Planck-Institut f\"ur Radioastronomie, Bonn, Germany
\item[$^{b}$] also at Kapteyn Institute, University of Groningen, Groningen, The Netherlands
\item[$^{c}$] School of Physics and Astronomy, University of Leeds, Leeds, United Kingdom
\item[$^{d}$] Fermi National Accelerator Laboratory, Fermilab, Batavia, IL, USA
\item[$^{e}$] Pennsylvania State University, University Park, PA, USA
\item[$^{f}$] Colorado State University, Fort Collins, CO, USA
\item[$^{g}$] Louisiana State University, Baton Rouge, LA, USA
\item[$^{h}$] now at Graduate School of Science, Osaka Metropolitan University, Osaka, Japan
\item[$^{i}$] Institut universitaire de France (IUF), France
\item[$^{j}$] now at Technische Universit\"at Dortmund and Ruhr-Universit\"at Bochum, Dortmund and Bochum, Germany
\end{description}

\section*{Acknowledgments}

\begin{sloppypar}
The successful installation, commissioning, and operation of the Pierre
Auger Observatory would not have been possible without the strong
commitment and effort from the technical and administrative staff in
Malarg\"ue. We are very grateful to the following agencies and
organizations for financial support:
\end{sloppypar}

\begin{sloppypar}
Argentina -- Comisi\'on Nacional de Energ\'\i{}a At\'omica; Agencia Nacional de
Promoci\'on Cient\'\i{}fica y Tecnol\'ogica (ANPCyT); Consejo Nacional de
Investigaciones Cient\'\i{}ficas y T\'ecnicas (CONICET); Gobierno de la
Provincia de Mendoza; Municipalidad de Malarg\"ue; NDM Holdings and Valle
Las Le\~nas; in gratitude for their continuing cooperation over land
access; Australia -- the Australian Research Council; Belgium -- Fonds
de la Recherche Scientifique (FNRS); Research Foundation Flanders (FWO),
Marie Curie Action of the European Union Grant No.~101107047; Brazil --
Conselho Nacional de Desenvolvimento Cient\'\i{}fico e Tecnol\'ogico (CNPq);
Financiadora de Estudos e Projetos (FINEP); Funda\c{c}\~ao de Amparo \`a
Pesquisa do Estado de Rio de Janeiro (FAPERJ); S\~ao Paulo Research
Foundation (FAPESP) Grants No.~2019/10151-2, No.~2010/07359-6 and
No.~1999/05404-3; Minist\'erio da Ci\^encia, Tecnologia, Inova\c{c}\~oes e
Comunica\c{c}\~oes (MCTIC); Czech Republic -- GACR 24-13049S, CAS LQ100102401,
MEYS LM2023032, CZ.02.1.01/0.0/0.0/16{\textunderscore}013/0001402,
CZ.02.1.01/0.0/0.0/18{\textunderscore}046/0016010 and
CZ.02.1.01/0.0/0.0/17{\textunderscore}049/0008422 and CZ.02.01.01/00/22{\textunderscore}008/0004632;
France -- Centre de Calcul IN2P3/CNRS; Centre National de la Recherche
Scientifique (CNRS); Conseil R\'egional Ile-de-France; D\'epartement
Physique Nucl\'eaire et Corpusculaire (PNC-IN2P3/CNRS); D\'epartement
Sciences de l'Univers (SDU-INSU/CNRS); Institut Lagrange de Paris (ILP)
Grant No.~LABEX ANR-10-LABX-63 within the Investissements d'Avenir
Programme Grant No.~ANR-11-IDEX-0004-02; Germany -- Bundesministerium
f\"ur Bildung und Forschung (BMBF); Deutsche Forschungsgemeinschaft (DFG);
Finanzministerium Baden-W\"urttemberg; Helmholtz Alliance for
Astroparticle Physics (HAP); Helmholtz-Gemeinschaft Deutscher
Forschungszentren (HGF); Ministerium f\"ur Kultur und Wissenschaft des
Landes Nordrhein-Westfalen; Ministerium f\"ur Wissenschaft, Forschung und
Kunst des Landes Baden-W\"urttemberg; Italy -- Istituto Nazionale di
Fisica Nucleare (INFN); Istituto Nazionale di Astrofisica (INAF);
Ministero dell'Universit\`a e della Ricerca (MUR); CETEMPS Center of
Excellence; Ministero degli Affari Esteri (MAE), ICSC Centro Nazionale
di Ricerca in High Performance Computing, Big Data and Quantum
Computing, funded by European Union NextGenerationEU, reference code
CN{\textunderscore}00000013; M\'exico -- Consejo Nacional de Ciencia y Tecnolog\'\i{}a
(CONACYT) No.~167733; Universidad Nacional Aut\'onoma de M\'exico (UNAM);
PAPIIT DGAPA-UNAM; The Netherlands -- Ministry of Education, Culture and
Science; Netherlands Organisation for Scientific Research (NWO); Dutch
national e-infrastructure with the support of SURF Cooperative; Poland
-- Ministry of Education and Science, grants No.~DIR/WK/2018/11 and
2022/WK/12; National Science Centre, grants No.~2016/22/M/ST9/00198,
2016/23/B/ST9/01635, 2020/39/B/ST9/01398, and 2022/45/B/ST9/02163;
Portugal -- Portuguese national funds and FEDER funds within Programa
Operacional Factores de Competitividade through Funda\c{c}\~ao para a Ci\^encia
e a Tecnologia (COMPETE); Romania -- Ministry of Research, Innovation
and Digitization, CNCS-UEFISCDI, contract no.~30N/2023 under Romanian
National Core Program LAPLAS VII, grant no.~PN 23 21 01 02 and project
number PN-III-P1-1.1-TE-2021-0924/TE57/2022, within PNCDI III; Slovenia
-- Slovenian Research Agency, grants P1-0031, P1-0385, I0-0033, N1-0111;
Spain -- Ministerio de Ciencia e Innovaci\'on/Agencia Estatal de
Investigaci\'on (PID2019-105544GB-I00, PID2022-140510NB-I00 and
RYC2019-027017-I), Xunta de Galicia (CIGUS Network of Research Centers,
Consolidaci\'on 2021 GRC GI-2033, ED431C-2021/22 and ED431F-2022/15),
Junta de Andaluc\'\i{}a (SOMM17/6104/UGR and P18-FR-4314), and the European
Union (Marie Sklodowska-Curie 101065027 and ERDF); USA -- Department of
Energy, Contracts No.~DE-AC02-07CH11359, No.~DE-FR02-04ER41300,
No.~DE-FG02-99ER41107 and No.~DE-SC0011689; National Science Foundation,
Grant No.~0450696, and NSF-2013199; The Grainger Foundation; Marie
Curie-IRSES/EPLANET; European Particle Physics Latin American Network;
and UNESCO.
\end{sloppypar}

}


\begin{thebibliography}{99}

\bibitem{masterclass} R. Sarmento et al. (The Pierre Auger Collaboration), PoS ICRC2023 (2023) 1611

\bibitem{opendata} Piera L. Ghia et al. (The Pierre Auger Collaboration), PoS ICRC2023 (2023) 1616

\bibitem{edumod} \url{https://mdetect.com.au/education/}

\bibitem{videobelgium} \url{https://www.iihe.ac.be/news/iihe-50th-anniversary-colloquium}

\bibitem{planetarium}
\url{https://halley.uis.edu.co/planetario/}

\bibitem{czechinterviews} \url{https://www.lip.pt/?section=press&page=news-details&id=1300}

\bibitem{corinne}
\url{https://www.sfp150ans.fr/evenements/conference-les-rayons-cosmiques-et-la-traque-des-plus-energetiques-a-lobservatoire-pierre-auger/}


\bibitem{Galati2024}
G. Galati, P. Paolucci, and F. Scianitti, 
\textit{“Art \& Science across Italy”: a journey between Science and Art in high schools}, PoS \textbf{ICHEP2024} (2024) 1170, \href{https://pos.sissa.it/476/1170/pdf}{doi:10.22323/1.476.1170}

\bibitem{palermo}
\url{https://www.instagram.com/p/DHLqpMZM8B7/?utm_source=ig_web_copy_link&igsh=MzRlODBiNWFlZA==}

\bibitem{fd}
J.~Abraham \textit{et al.}, \emph{The fluorescence detector of the Pierre Auger Observatory}, 
\emph{Nucl.\ Instrum.\ Meth.\ A} \textbf{620} (2010) 227--251, \href{https://doi.org/10.1016/j.nima.2010.04.023}{doi:10.1016/j.nima.2010.04.023}.

\bibitem{polandfilm} \url{https://youtu.be/1QO41a8A-hQ}

\bibitem{astrofest}
\url{https://stiintasitehnica.com/astrofest-2023-20-mai-parcul-crangasi/}

\bibitem{atmosfericos} \url{https://www.crtvg.es/tvg/a-carta/programa-237-6336542?t=828}


\bibitem{womenscience} \url{https://www.auger.org/outreach/equal-opportunities/women-in-auger}

\bibitem{arizuspace} \url{https://mendoza.tur.ar/travel-directory/espacio-arizu/}

\bibitem{iauworkinggroup} \url{https://iau-oao.nao.ac.jp/iau-inclusion/home/#:~:text=Outlining%20and%20prioritizing%20diversity%20this,cultural%2C%20racial%2C%20language%20and%20religious}

\bibitem{Casado1} Casado, J., García, B., Gandhi, P., Díaz-Merced, W. (2022). A New Approach to Sonification of Astrophysical Data: The User Centred Design of SonoUno. American Journal of Astronomy and Astrophysics, 9(4), 42-51. https://doi.org/10.11648/j.ajaa.20210904.11


\bibitem{augerenfoco} \url{https://visitantes.auger.org.ar/index.php/auger-en-foco/}


\end{thebibliography}
\end{document}